\documentclass[12pt]{article}

\input{tcilatex}

\begin{document}

\bigskip

\bigskip\ 

\begin{center}
\textbf{EQUIVALENCE BETWEEN VARIOUS VERSIONS OF THE}

\smallskip\ 

\textbf{SELF-DUAL ACTION OF THE ASHTEKAR FORMALISM}

\bigskip\ 

\smallskip\ 

J. A. Nieto\footnote[1]{%
nieto@uas.uasnet.mx}

\smallskip\ 

\textit{Facultad de Ciencias F\'{\i}sico-Matem\'{a}ticas de la Universidad
Aut\'{o}noma}

\textit{de Sinaloa, 80010, Culiac\'{a}n Sinaloa, M\'{e}xico}

\bigskip\ 

\bigskip\ 

\textbf{Abstract}
\end{center}

Different aspects of the self-dual (anti-self-dual) action of the Ashtekar
canonical formalism are discussed. In particular, we study the equivalences
and differences between the various versions of such an action. Our analysis
may be useful for the development of Ashtekar formalism in eight dimensions.

\bigskip\ 

\bigskip\ 

\bigskip\ 

\bigskip\ 

\bigskip\ 

\bigskip\ 

\bigskip\ 

Pacs numbers: 04.60.-m, 04.65.+e, 11.15.-q, 11.30.Ly

April, 2004

\newpage

Recently, Ashtekar and Lewandowski [1] reported a pedagogical presentation
of loop quantum gravity. In particular, in its review article they consider
the action

\begin{equation}
S_{1}=\frac{1}{4}\int_{\mathcal{M}}\epsilon _{IJKL}e^{I}\wedge e^{J}\wedge
\Omega ^{KL}-\frac{1}{2\gamma }\int_{\mathcal{M}}e^{I}\wedge e^{J}\wedge
\Omega _{IJ}.  \tag{1}
\end{equation}%
as the starting point in the program of non-perturbative quantum gravity.
Here, we closely follow the notation in Ref. [1]. $\mathcal{M}$ denotes a
four dimensional spacetime manifold, $e^{I}$ are co-tetrads defined in $T_{x}%
\mathcal{M}$ for each $x\in \mathcal{M}$, $\epsilon _{IJKL}$ is a completely
antisymmetric tensor compatible with the invariant metric $\eta
_{IJ}=diag(\pm 1,1,1,1)$ of the group $SO(1,3)$ or $SO(4)$ and

\begin{equation}
\Omega \equiv d\omega +\omega \wedge \omega .  \tag{2}
\end{equation}%
The quantity $\gamma $ is a number called the Barebero-Immirzi parameter. If
the signature of $\eta _{IJ}$ is $0+4$ the interesting geometrical structure
arises when $\gamma =\pm 1$, while if the signature of $\eta _{IJ}$ is
Lorenziana one has $\gamma =\pm i.$ Just for simplicity, we shall focus in
the case $\gamma =i$. Nevertheless, most of our computations also apply to
the cases $\gamma =-i$ and $\gamma =\pm 1$.

The abstract notation in (1) is, in some sense, elegant but some times for
computations is not so practical as a tensorial notation. For this reason,
in order to clarify some aspects of the action (1) it becomes convenient to
rewrite (1) in a tensorial notation. Since $e^{I}$ and $\Omega ^{Kl}$ are
one-form and two-form respectively we have $e^{I}=e_{\mu }^{I}dx^{\mu }$ and 
$\Omega ^{KL}=\frac{1}{2}\Omega _{\mu \nu }^{KL}dx^{\mu }\wedge dx^{\nu }$,
where

\begin{equation}
\Omega _{\mu \nu }^{Kl}=\partial _{\mu }\omega _{\nu }^{KL}-\partial _{\nu
}\omega _{\mu }^{KL}+\omega _{\mu }^{KS}\omega _{\nu S}^{L}-\omega _{\mu
}^{LS}\omega _{\nu S}^{K}.  \tag{3}
\end{equation}

Let write the action (1) in terms of $e_{\mu }^{I}$ and $\Omega _{\mu \nu
}^{KL}$. One has

\begin{equation}
S_{1}=\frac{1}{8}\int d^{4}x\epsilon ^{\mu \nu \alpha \beta }\epsilon
_{IJKL}e_{\mu }^{I}e_{\nu }^{J}\Omega _{\alpha \beta }^{KL}+\frac{i}{8}\int
d^{4}x\epsilon ^{\mu \nu \alpha \beta }\delta _{IJRS}e_{\mu }^{I}e_{\nu
}^{J}\Omega _{\alpha \beta }^{RS}.  \tag{4}
\end{equation}%
Here, we used the notation $\delta _{IJRS}=\eta _{IR}\eta _{JS}-\eta
_{IS}\eta _{JR}$ and the fact that

\begin{equation}
\int d^{4}x\epsilon ^{\mu \nu \alpha \beta }=\int_{\mathcal{M}}dx^{\mu
}\wedge dx^{\nu }\wedge dx^{\alpha }\wedge dx^{\beta }.  \tag{5}
\end{equation}%
Now, considering that

\begin{equation}
\epsilon _{IJKL}\epsilon _{RS}^{KL}=-2\delta _{IJRS}  \tag{6}
\end{equation}%
and by using the notation

\begin{equation}
\Sigma _{\mu \nu }^{IJ}\equiv e_{\mu }^{I}e_{\nu }^{J}-e_{\mu }^{J}e_{\nu
}^{I}  \tag{7}
\end{equation}%
one sees that the action (4) can also be rewritten as

\begin{equation}
S_{1}=\frac{1}{16}\int d^{4}x\epsilon ^{\mu \nu \alpha \beta }\epsilon
_{IJKL}\Sigma _{\mu \nu }^{IJ}\Omega _{\alpha \beta }^{KL}-\frac{i}{32}\int
d^{4}x\epsilon ^{\mu \nu \alpha \beta }\epsilon _{IJKL}\epsilon
_{RS}^{KL}\Sigma _{\mu \nu }^{IJ}\Omega _{\alpha \beta }^{RS}.  \tag{8}
\end{equation}%
It is straightforward to see that this action is equivalent to

\begin{equation}
S_{1}=\frac{1}{8}\int d^{4}x\epsilon ^{\mu \nu \alpha \beta }\Sigma _{\mu
\nu }^{IJ+}\Omega _{\alpha \beta }^{KL}\epsilon _{IJKL},  \tag{9}
\end{equation}%
where

\begin{equation}
^{+}\Omega _{\alpha \beta }^{KL}\equiv \frac{1}{2}(\Omega _{\alpha \beta
}^{KL}-\frac{i}{2}\epsilon _{RS}^{KL}\Omega _{\alpha \beta }^{RS}).  \tag{10}
\end{equation}%
Thus, we have proved step by step that (1) is equivalent to (9).

It turns out that an alternative, but equivalent, way to write \ (9) is

\begin{equation}
S_{1}=\int d^{4}xee_{K}^{\alpha }e_{L}^{\beta +}\Omega _{\alpha \beta }^{KL}.
\tag{11}
\end{equation}%
Here, we used (7) and the property that $\epsilon ^{\mu \nu \alpha \beta
}e_{\mu }^{I}e_{\nu }^{J}\epsilon _{IJKL}=2e\Sigma _{KL}^{\alpha \beta }$,
with $e=\det (e_{\mu }^{I}).$ One recognizes in the expression (11) the
action proposed by Jacobson and Smolin [2] and Samuel [3]. Therefore, the
action (1) called in Ref. [1] the Holst action [4] is just the same as the
one proposed by Jacobson-Smolin-Samuel for $\gamma =i$.

An important aspect of (11) or (9) is that it can be reduced to the real
Einstein-Hilbert action

\begin{equation}
S_{EH}=\frac{1}{2}\int d^{4}xee_{K}^{\alpha }e_{L}^{\beta }\Omega _{\alpha
\beta }^{KL}  \tag{12}
\end{equation}%
or

\begin{equation}
S_{EH}=\frac{1}{16}\int d^{4}x\epsilon ^{\mu \nu \alpha \beta }\Sigma _{\mu
\nu }^{IJ}\Omega _{\alpha \beta }^{KL}\epsilon _{IJKL}.  \tag{13}
\end{equation}%
This result holds by the following reasons. First of all, observe that $%
^{+}\Omega _{\alpha \beta }^{KL}$ is self-dual,

\begin{equation}
^{\ast +}\Omega _{\alpha \beta }^{KL}\equiv i^{+}\Omega _{\alpha \beta
}^{KL}.  \tag{14}
\end{equation}%
Here, we used the definition $^{\ast +}\Omega _{\alpha \beta }^{KL}=\frac{1}{%
2}\epsilon _{RS}^{KL+}\Omega _{\alpha \beta }^{RS}$. Therefore, $^{+}\Omega
_{\alpha \beta }^{KL}$ keeps only the self-dual part of the real curvature $%
\Omega _{\alpha \beta }^{KL}$. It turns out that, as Jacobson and Smolin
emphasize [23], in four dimensions one has the algebra isomorphism $%
so(1,3)=su(2)\times su(2)$ and consequently the self-dual curvature $%
^{+}\Omega _{\alpha \beta }^{KL}$ can be combined with the anti-self-dual
curvature%
\begin{equation}
^{-}\Omega _{\alpha \beta }^{KL}=\frac{1}{2}(\Omega _{\alpha \beta }^{KL}+%
\frac{i}{2}\epsilon _{RS}^{KL}\Omega _{\alpha \beta }^{RS})  \tag{15}
\end{equation}%
in the form

\begin{equation}
\Omega _{\alpha \beta }^{KL}(^{+}\omega +^{-}\omega )=\Omega _{\alpha \beta
}^{KL}(^{+}\omega )+\Omega _{\alpha \beta }^{KL}(^{-}\omega ).  \tag{16}
\end{equation}%
Here, we used the fact that $\Omega _{\alpha \beta }^{KL}(^{\pm }\omega
)=^{\pm }\Omega _{\alpha \beta }^{KL}$, where

\begin{equation}
^{\pm }\omega _{\alpha }^{KL}=\frac{1}{2}(\omega _{\alpha }^{KL}\mp \frac{i}{%
2}\epsilon _{RS}^{KL}\omega _{\alpha }^{RS}).  \tag{17}
\end{equation}%
Considering these results one finds that the equation of motion derived from
the action (11) under variations with respect to $^{+}\omega $ leads to the
cyclic Bianchi identity for $\Omega _{\alpha \beta }^{KL}(e)$ and therefore
the second term in (8), or (4), vanishes identically (see Ref. [2] for more
details).

A variant of the action (9) is provided by the action

\begin{equation}
S_{2}=\frac{1}{8}\int d^{4}x\epsilon ^{\mu \nu \alpha \beta +}\Sigma _{\mu
\nu }^{IJ+}\Omega _{\alpha \beta }^{KL}\delta _{IJKL},  \tag{18}
\end{equation}%
where $^{\pm }\Sigma _{\mu \nu }^{IJ}$ is the (anti-)self-dual part of $%
\Sigma _{\mu \nu }^{IJ}$:

\begin{equation}
^{\pm }\Sigma _{\mu \nu }^{IJ}=\frac{1}{2}(\Sigma _{\mu \nu }^{IJ}\mp \frac{i%
}{2}\epsilon _{RS}^{KL+}\Sigma _{\mu \nu }^{RS}).  \tag{19}
\end{equation}%
In abstract notation the action (18) becomes

\begin{equation}
S_{2}=\int_{\mathcal{M}}\Sigma _{(+)}^{IJ}\wedge \Omega _{IJ}^{(+)}. 
\tag{20}
\end{equation}%
Here, $\Sigma _{(+)}^{IJ}=\frac{1}{2}^{+}\Sigma _{\mu \nu }^{IJ}dx^{\mu
}\wedge dx^{\nu }$ and $\Omega _{IJ}^{(+)}=\frac{1}{2}^{+}\Omega _{\alpha
\beta }^{KL}dx^{\alpha }\wedge dx^{\beta }$. It turns out that the action
(20), called $S_{(H)}$ in Ref. [1] (see expression (2.9) in Ref. [1]), also
plays an essential role in the canonical quantization of gravity in four
dimensions (see [1] and Refs. therein).

In this work, we shall show that the actions (9) and (18) (or equivalent (1)
and (20)) can be considered as part of the action

\begin{equation}
S_{3}^{(+)}=\frac{1}{4}\int_{\mathcal{M}}\mathcal{F}^{(+)IJ}\wedge \mathcal{F%
}^{(+)KL}\epsilon _{IJKL},  \tag{21}
\end{equation}%
which was proposed in Ref. [5] and generalized to the supersymmetric case in
Ref. [6]. Here, the de Sitter curvature $\mathcal{F}^{IJ}=\frac{1}{2}%
\mathcal{F}_{\mu \nu }^{IJ}dx^{\mu }\wedge dx^{\nu }$ is defined as

\begin{equation}
\mathcal{F}_{\mu \nu }^{IJ}=\Omega _{\mu \nu }^{IJ}+\Sigma _{\mu \nu }^{IJ}.
\tag{22}
\end{equation}

First , observe the action (21) can be rewritten as

\begin{equation}
S_{3}=\frac{1}{16}\int d^{4}x\epsilon ^{\mu \nu \alpha \beta +}\mathcal{F}%
_{\mu \nu }^{IJ+}\mathcal{F}_{\alpha \beta }^{KL}\epsilon _{IJKL}.  \tag{23}
\end{equation}%
Let us now write $^{+}\mathcal{F}_{\mu \nu }^{IJ}$ in the form%
\begin{equation}
^{+}\mathcal{F}_{\mu \nu }^{IJ}=\frac{1}{2}^{+}B_{KL}^{IJ}\mathcal{F}_{\mu
\nu }^{KL},  \tag{24}
\end{equation}%
where

\begin{equation}
^{\pm }B_{KL}^{IJ}=\frac{1}{2}(\delta _{KL}^{IJ}\mp i\epsilon _{KL}^{IJ}). 
\tag{25}
\end{equation}%
By straightforward computation we find that the projector $^{\pm
}B_{KL}^{IJ} $ has the following interesting properties:

\begin{equation}
^{\pm }B_{MN}^{IJ}\epsilon _{IJKL}=\pm 2i^{\pm }B_{MNKL},  \tag{26}
\end{equation}

\begin{equation}
^{\pm }B_{MN}^{IJ}\delta _{IJKL}=2^{\pm }B_{MNKL},  \tag{27}
\end{equation}

\begin{equation}
^{\pm }B_{MN}^{IJ\pm }B_{RS}^{KL}\epsilon _{IJKL}=\pm 4i^{\pm }B_{MNRS} 
\tag{28}
\end{equation}%
and

\begin{equation}
^{\pm }B_{MN}^{IJ\pm }B_{RS}^{KL}\delta _{IJKL}=4^{\pm }B_{MNRS}.  \tag{29}
\end{equation}%
Moreover, (28) and (29) can be added to give

\begin{equation}
^{\pm }B_{MN}^{IJ\pm }B_{RS}^{KL\pm }B_{IJKL}=4^{\pm }B_{MNRS}.  \tag{30}
\end{equation}%
Now, consider the alternative action

\begin{equation}
S_{4}=\frac{1}{16}\int d^{4}x\epsilon ^{\mu \nu \alpha \beta +}\mathcal{F}%
_{\mu \nu }^{IJ+}\mathcal{F}_{\alpha \beta }^{KL}\delta _{IJKL}.  \tag{31}
\end{equation}%
By using the relations (28) and (29) one discovers that

\begin{equation}
S_{3}=+iS_{4}.  \tag{32}
\end{equation}%
Therefore, up to the complex numerical factor $i$ the actions $S_{3}$ and $%
S_{4}$ are equal. Thus, one can use either $S_{3}$ or $S_{4}$ to get the
same gravitational information.

Let us now focus on the action $S_{4}$. Using (22) one sees that $S_{4}$
leads to

\begin{equation}
\begin{array}{c}
S_{4}=\frac{1}{16}\int d^{4}x\epsilon ^{\mu \nu \alpha \beta +}\Omega _{\mu
\nu }^{IJ+}\Omega _{\alpha \beta }^{KL}\delta _{IJKL}+\frac{1}{8}\int
d^{4}x\epsilon ^{\mu \nu \alpha \beta +}\Sigma _{\mu \nu }^{IJ+}\Omega
_{\alpha \beta }^{KL}\delta _{IJKL} \\ 
\\ 
+\frac{1}{16}\int d^{4}x\epsilon ^{\mu \nu \alpha \beta +}\Sigma _{\mu \nu
}^{IJ+}\Sigma _{\alpha \beta }^{KL}\delta _{IJKL}.%
\end{array}
\tag{33}
\end{equation}%
The first term in (33) corresponds to the complex sum of Euler and
Prontrjagin topological invariants. The last term refers to the cosmological
constant term. While the second term corresponds to the tensorial version of
the action (20) (see action (18)). Therefore, we have proved that the action
(18) is obtained from (31) when one dropps from $S_{4}$ the Euler and
Prontrjagin topological invariants and the cosmological term. But, since $%
S_{4}$ is classically equivalent to the action $S_{3}$ this also proves that
the action (18) or the action (20) are contained in the action $S_{3}$.

Using the relations (28) and (29) it can be shown that action $S_{2}$, given
in (18) (or (20)), is equivalent to

\begin{equation}
S_{5}=\frac{1}{8}\int d^{4}x\epsilon ^{\mu \nu \alpha \beta +}\Sigma _{\mu
\nu }^{IJ+}\Omega _{\alpha \beta }^{KL}\epsilon _{IJKL}.  \tag{34}
\end{equation}%
In fact, we find that $S_{5}=iS_{2}.$ Moreover, using the relations
(26)-(29) one finds that $S_{5}$ can be written in a variety of equivalent
ways:

\begin{equation}
S_{6}=\frac{i}{8}\int d^{4}x\epsilon ^{\mu \nu \alpha \beta }\Sigma _{\mu
\nu }^{IJ}\Omega _{\alpha \beta }^{KL+}B_{IJKL},  \tag{35}
\end{equation}

\begin{equation}
S_{7}=\frac{i}{8}\int d^{4}x\epsilon ^{\mu \nu \alpha \beta }\Sigma _{\mu
\nu }^{IJ+}\Omega _{\alpha \beta }^{KL+}B_{IJKL},  \tag{36}
\end{equation}

\begin{equation}
S_{8}=\frac{i}{4}\int d^{4}x\epsilon ^{\mu \nu \alpha \beta }\Sigma _{\mu
\nu }^{IJ+}\Omega _{\alpha \beta IJ},  \tag{37}
\end{equation}

\begin{equation}
S_{9}=\frac{1}{8}\int d^{4}x\epsilon ^{\mu \nu \alpha \beta }\Sigma _{\mu
\nu }^{IJ+}\Omega _{\alpha \beta }^{KL}\epsilon _{IJKL},  \tag{38}
\end{equation}%
In fact, we find that $S_{5}=S_{6}=S_{7}=S_{8}=S_{9}$. In particular one
observes that $S_{1}=S_{9}$ and therefore one discovers that $S_{1}=iS_{2}$.

Just for completeness let us write the actions (34)-(38) in abstract
notation:

\begin{equation}
S_{5}=\frac{1}{2}\int_{\mathcal{M}}\Sigma _{(+)}^{IJ}\wedge \Omega
^{(+)KL}\epsilon _{IJKL},  \tag{39}
\end{equation}

\begin{equation}
S_{6}=\frac{i}{2}\int_{\mathcal{M}}\Sigma ^{IJ}\wedge \Omega ^{KL+}B_{IJKL},
\tag{40}
\end{equation}

\begin{equation}
S_{7}=\frac{i}{2}\int_{\mathcal{M}}\Sigma ^{IJ}\wedge \Omega
^{(+)KL+}B_{IJKL},  \tag{41}
\end{equation}

\begin{equation}
S_{8}=i\int_{\mathcal{M}}\Sigma ^{IJ}\wedge \Omega _{IJ}^{(+)},  \tag{42}
\end{equation}

\begin{equation}
S_{9}=\frac{1}{2}\int_{\mathcal{M}}\Sigma ^{IJ}\wedge \Omega
^{(+)KL}\epsilon _{IJKL},  \tag{43}
\end{equation}

If instead of self-dual sector $^{+}\mathcal{F}^{IJ}$ one considers the
anti-self-dual sector $^{-}\mathcal{F}^{IJ}$ of $\mathcal{F}^{IJ}$ one
obtains the analogue action to $S_{3}^{(+)},$ namely

\begin{equation}
S_{3}^{(-)}=\frac{1}{4}\int_{\mathcal{M}}\mathcal{F}^{(-)IJ}\wedge \mathcal{F%
}^{(-)KL}\epsilon _{IJKL}.  \tag{44}
\end{equation}%
Following similar steps as in the case of $S_{3}^{(+)}$ one may obtain from $%
S_{3}^{(-)}$ all the corresponding equivalent actions $S_{5},...,S_{9}$
given in (39)-(43) but with the sign $(+)$ replaced by the sign $(-).$

Summarizing, we have proved that the actions (1) and (20), reported in the
review [1], are particular cases of the more general action $S_{3}^{(+)}$
(or $S_{3}^{(-)}$). Specifically, (1) and (20) are obtained when one
discards from $S_{3}^{(+)}$ the Euler and Pontrjagin topological invariants
and the cosmological constant term.

From (16) and (22) one sees that $\mathcal{F}^{IJ}=\mathcal{F}^{(+)IJ}+%
\mathcal{F}^{(-)IJ}$ and consequently one finds

\begin{equation}
\begin{array}{c}
S_{MM}=\frac{1}{4}\int_{\mathcal{M}}\mathcal{F}^{IJ}\wedge \mathcal{F}%
^{KL}\epsilon _{IJKL}=\frac{1}{4}\int_{\mathcal{M}}\mathcal{F}^{(+)IJ}\wedge 
\mathcal{F}^{(+)KL}\epsilon _{IJKL} \\ 
\\ 
+\frac{1}{4}\int_{\mathcal{M}}\mathcal{F}^{(-)IJ}\wedge \mathcal{F}%
^{(-)KL}\epsilon _{IJKL}+\frac{1}{2}\int_{\mathcal{M}}\mathcal{F}%
^{(+)IJ}\wedge \mathcal{F}^{(-)KL}\epsilon _{IJKL}.%
\end{array}
\tag{45}
\end{equation}%
One recognizes the action $S_{MM}$ as the usual MacDowell-Mansouri action
[7] (see also Ref. [8]) which can be derived by breaking the De Sitter gauge
group $SO(1,4)$ (or anti-De Sitter $SO(2,3)$) to the smaller group $SO(1,3)$%
. In fact, in the MacDowell-Mansouri theory the connection $\omega ^{IJ}$
associated with the group $SO(1,3)$ and the tetrad $e^{I}$ arise after
breaking the original De Sitter connection $\omega ^{\hat{I}\hat{J}}$ where $%
\hat{I}$ and $\hat{J}$ are group indices of the De Sitter group $SO(1,4)$
(or anti-De Sitter group $SO(2,3)$). Consequently, $S_{3}^{(+)}$ (or $%
S_{3}^{(-)}$) carries the self-dual sector of the De Sitter gauge group
property of $S_{MM}$.

On purpose to study $S-$duality for gravity Garc\'{\i}a-Compe\'{a}n \textit{%
et al} [8] modified the action (45) in the form

\begin{equation}
\mathcal{S}_{3}=\frac{^{+}\tau }{4}\int_{\mathcal{M}}\mathcal{F}%
^{(+)IJ}\wedge \mathcal{F}^{(+)KL}\epsilon _{IJKL}+\frac{^{-}\tau }{4}\int_{%
\mathcal{M}}\mathcal{F}^{(-)IJ}\wedge \mathcal{F}^{(-)KL}\epsilon _{IJKL}. 
\tag{46}
\end{equation}%
where $^{+}\tau $ and $^{-}\tau $ are constant coupling parameters. Indeed,
the action (46) is the bosonic sector a more general supersymmetric action
(see Ref. [9] for details).

Other generalization of $S_{3}^{(+)}$ seems to lack of the attractive gauge
properties contained in MacDowell-Mansouri procedure. In particular, by
combining duality in the spacetime indices and the group indices Soo [10]
extended the action $S_{3}^{(+)}$ to a positive definite action. Montesinos
[11] added to (1) the Euler and Pontrjagin topological invariants with
constant factor parameters. Obukhov and Hehl [12] analized also a number of
possibilities by combining dualities in the spacetime indices and the group
indices. However, since a duality associated with the spacetime indices
requires a metric these extensions of $S_{3}^{(+)}$ can not be considered as
genuine gauge theories in the sense of the MacDowell-Mansouri theory.

Besides the important gauge properties of the action $S_{3}^{(+)}$ one may
become interested in this action because of its closeness to topological
features. In fact, as has been pointed out [5], if instead of the fully
antisymmetric symbol $\epsilon _{IJKL}$ one uses the Killing metric
associated with the De Sitter group $SO(1,4)$ (or $SO(2,3)$) the action $%
S_{3}^{(+)}$ becomes the second Chern class. This means that the action $%
S_{3}^{(+)}$ which leads to the Ashtekar formalism is closely related to the
Chern-Simons action. Thus, a canonical quantization of $S_{3}^{(+)}$ may
lead to the intriguing result that an `almost' a Chern-Simons action, as is
the case of $S_{3}^{(+)}$, leads to physical states which are exponential of
the Chern-Simons action which are predicted by the quantum general
relativity of the canonical Ashtekar formalism (see [13]-[14]).

Finally, the present work may be useful in the recent proposal [15] of
extending the Ashtekar formalism to eight dimensions. It turns out that, by
using an octonionic structure [16]-[17], in the Ref. [15] the analogue of
the action $S_{3}^{(+)}$ was proposed in a spacetime of signature $1+7$.
Thus, a revisited analysis of the action $S_{3}^{(+)},$ as presented in this
work, seems to be a necessary step for further development in such an eight
dimensional program.

\end{document}